\documentclass[letterpaper, 10 pt, conference]{ieeeconf}  

\IEEEoverridecommandlockouts                              

\usepackage{graphicx}
\usepackage{amsmath}
\title{\LARGE \bf
A New Model of Thermal Propagation in Human Tissue by Using HIFU Application   
}

\author{Saeed Reza Hajian $^{1}$, Ali Abbaspour Tehrani Fard $^{2}$, Majid Pouladian $^{3,*}$ \\ and Gholam Reza Hemmasi $^{4}$
\thanks{$^{1}$ Medical Radiation Eng. Department, Engineering Faculty, Science and Research Branch, Islamic Azad University, Tehran, Iran}
\thanks{$^{2}$ Assistant Professor, Department of Electrical Engineering, Sharif University of Technology, Tehran, Iran}
\thanks{$^{3}$ Chief of Biomedical Eng. Faculty, Science and Research Branch, Islamic Azad, University, Tehran, Iran
	}
\thanks{$^{4}$ Assistant professor, Research Centre of Gastroenterology \& Hepatology of Firouzgar Hospital, Iran University of Medical Science, Tehran, Iran}
\thanks{$^{*}$ Corresponding Author: Majid Pouladian, Head of Biomedical  Engineering Faculty, Science and Research Branch, Islamic Azad University,Tehran (Iran), Tel: +989121390336; Fax: +982122834605, \tt\small Pouladian@srbiau.ac.ir}
\thanks{\textbf{Keywords: \textit{Thermal Index, HIFU (High Intensity Focus Ultrasound), Mechanical Modelling, Extra corporeal, Intensity, Ultrasound Beam, Mechanical Wave}} } 
}

\begin{document}

\maketitle
\thispagestyle{empty}
\pagestyle{empty}

\begin{abstract}

In outside the body HIFU treatment that focused ultrasound beams hit severely with cancer tissue layer especially the soft one, at the time of passage of the body different layers as long as they want to reach tumor, put their own way components under mechanical and even thermal influence and they can cause skin lesions. To reduce this effect a specific mechanical model can be used that means body tissue is considered as a mechanical model, it is affected when passing sound mechanical waves through it and each layer has an average heat. Gradually sound intensity decreases through every layer passage, finally in one direction a decreased intensity sound reach tumor tissue. If sound propagated directions increase, countless waves with decreased intensity are gathered upon the tumor tissue that causes a lot of heat focus on tumor tissue. Depending on the kind and mechanical properties of the tissue, intensity of each sound wave when it passes through tissue can be controlled to reduce damages outside the tumor tissue.

\end{abstract}

\section{Introduction}
Today, ultrasound beams apart from diagnosis and ultrasound imaging is used in treatment process such as the treatment of cancers (HIFU), stone crushing in the body, outside the body, rehabilitation and injection and detection of hematological parameters. Cancers treatment is one of ultrasound uses to which is done inside the body and outside the body. Some external damages and skin lesions have been reported in outside the body therapy methods which are used more in treatment of abdominal soft tissue cancers. Many of the articles have worked on methods of heat gaining during the therapy with HIFU but examining method based on tissue mechanical model is a new approach. \cite{aghayan2013inverse} 
Today, several people have worked on thermal effects during treatment. In one of these reports, the thermal effect on the tissue and its cooling system is considered and it is tried to obtain thermal coefficients using reverse engineering in solving bioheat equation that don’t harm tissue. Of course this action is considered in time process and in hyperthermia treatment but tissue is severed in HIFU and physiotherapy practice isn’t done.
\\
In discussing HIFU and its thermal effects in the treatment process, when the ultrasound waves are radiated tissue, because changing in some tissue thermal parameters and by simulation of KZK equations using acoustic wave’s field and coefficients of tissue, heat distribution is calculated at the time of treatment. The dimensions of the transducer, the distance from the tissue and radiation time is also effective in heat distribution along with the tissue coefficients and tissue specific heat coefficient.  \cite{guntur2015influence} \cite{Sarraf_2016} \cite{grady2016age}
In HIFU treatment, laser radiation could be used along with the sound waves. Since laser waves have wavelength and every body tissue absorbs a wavelength can be absorbed. In fact they select that tissue. When the laser is absorbed in the tissue, causes temperature raise of the tissue. In the meantime, ultrasound waves can be radiated to the desired tissue. Being these together makes the treatment more precise and at the same time increase of the tissue heat. However in this manner when the depth is greater laser can’t have a major influence? \cite{kim2014dual}. 
\\
Usually bio heat equations are used to simulate the heat. Sound waves are propagated with a wave equation. Propagated wave form, its distribution manner has an impact on the KZK equation.
Diffraction occurs when the sound enters the tissue. Existence of this phenomenon causes waves are constantly diffracting when they come out of the piezoelectric cell until they reach the target tumor tissue. Number of piezoelectric cells, transducer diameter, and distance from the tissue, the tissue thickness and the overall angle of diffraction causes a heat and pressure at the center tissue. Different methods except ionizing methods such as microwave, radiofrequency, laser, optical and ultrasound methods are used for the treatment of soft tissue cancers. \cite{wang2016simulation} 
\\
In all of these methods, it is important to detect heat amount in the treatment process. MRI imaging techniques can be used for heat tracing. Treatment process heat tracking with this method, is called MR thermometry. MRI in this way is used in many medical treatments for examining cancer treatment improving and tracing accuracy of treatment and comparison is made. \cite{kim2014mr} \cite{sarraf2014brain} \cite{sarraf2014mathematical}
Piezoelectric cells that are used for treatment in HIFU, are in a spherical dish and make focal intensified waves in the tumor spot, they can be used to make a tissue slow or treat it in non-invasive manner. Geometry form and dimensions of the transducer, gender like PZT and type of piezoelectric cells arrangement, and piezoelectric cells acoustic impedance are important in type of generated heat. R. Martinez, A., Vera, L. and colleagues marked the piezoelectric cells in this way. They could obtain best type of piezoelectric cell and its arrangement based on frequency and based on a finite element method based on MATLAB simulation software. This examining method causes reducing the least heat in the treatment process that is performed in the piezoelectric cell and treatment process is done more efficiently. \cite{martinez2014heat} \cite{sarraf2016deepad}
\\
At the time of treatment with HIFU the dose term is used at the time of energy receiving in the tumor area. M Costaand his colleges at the time of treatment  with HIFU used optical CT device and imaging simultaneously with a volumetric chemical dosimeter called PRESAGE®Which is made of polyurethane and at the time of using sound waves changes color according to heat in the focal point based on optical imaging for heat calibration of HIFU device. \cite{costa2015presage} \cite{sarraf2016robust}
Using gel phantom with entering some bubbles in it and applying HIFU waves, heat degree varies despite existence of some bubbles. If bubbles are injected vibrationally, convert ultrasound kinetic energy into thermal energy. Radial movements and changes of HIFU waves in this phantom are simulated numerically and then are compared with the actual value and the impact of micro-bubbles is reviewed in the process. \cite{tamura2014visualizations} 
\\
The overall structure of this article is to explain design model components in Section 2, mechanical model components of the tissue in Section 3, results and laboratory works in Section 4, the benefits of this method in Section 5 and overall results of the study in 6.

\section{STATEMENT OF THE PROBLEM}
Since in treatment with HIFU mechanical beam is used and treatment is done with two ways of inside the body and outside the body, in outside the body treatment methods, if tumor is in body depth, intensified sound beam should pass soft tissues before hitting tumor. Because of being mechanical, wave influences on healthy soft tissues and even in some cases causes dermal damages. Producer Company named China HIFU reported some skin lesions in treatment of soft tissue tumors. Despite inventing MR-HIFU systems, led us toward using a mechanical model and using tissue binary feature that tissue parameters can be extracted using elastography or lab, made entering sound to tissue mechanical model and obtained pressure and heat in each layer through Simulink MATLAB.  If the temperature or pressure in any direction is to the extent that cause skin lesions, in the same direction is extracted with respect to tissue parameters and in biophysical cases that the temperature degree  should not exceed 2 $^{\circ}$ C  in order to damage not occur, sound wave intensity from transducer in the same direction can be changed. 
The model creating and modeling manner in the software environment is a unique method that nowhere pressure and heat extracted from HIFU had not been discussed in this way. However, this amount of extracted heat and pressure should be compared with practical example that unfortunately due to lack of HIFU device in Iran, the extracted cases and done with HIFU simulation  are considered by the FDA as an official reference of  data comparison and being standard. All the variables, variable with time in the project are not considered, and even the blood supply system that is a cooling parameter at the time of treatment with HIFU, is not considered. Hope that in the future with the acquisition of equipment and practical tests with completing the simulation and adding mentioned parameters using Simulink Sim mechanical environment animation; works more like the reality are done.
In the project using a sheep kidney, measured and divided in equal parts in lab using mechanical parameters measurement device,  obtained parameters were measured and entered simulator environment and with applying a HIFU wave in one direction, resulting heat on a kidney tissue that is in phantom is obtained.

\section{Material and Methods}
In the model where is based on three elements of mass, spring and damper using Simulink Toolbox   MATLAB software version2015a has been used. In HIFU process that mechanical sound waves are applied to the tissue based on ultrasound focused and intensified waves, if the tissue is assumed to be a mechanical tissue, it can be considered to be comprised of three members of mass, spring and damper. Damper and spring values according to tissue coefficients are derived from the following equation:
\begin{equation}\label{eq1}
F\_Spring=k.u
\end{equation} 
\begin{equation}\label{eq2}
F\_Dashpot= \eta  . (u)^{.} 
\end{equation}
In these equations, F is applied force to the spring or damper, k, spring rate and η is the coefficient of damper viscosity. 
In physics or tissue lab, tissue values can be obtained, obtaining viscosity and springy values of tissue based on simulation in MATLAB software, amount of heat in each layer of tissue is obtained. If the type of tissue, for example, is supposed to be a kidney, knowing it’s acoustic impedance degree and putting the speed of sound in soft tissue and kidney tissue acoustic impedance whose values are respectively the values in Table \ref{tabel1} \cite{hedrick2005ultrasound} \cite{saverino2016associative}. Using the HIFU simulator software [10] that has been mentioned in FDA site and is to simulate HIFU, the pressure and temperature at each layer of tissue is obtained and compared.
\begin{table}[htb]
	\centering
	\caption{Acoustic impedances of different body tissues and organs by Food and Drug Administration}
	\label{tabel1}
	\begin{tabular}{ll}
		\hline
		Body tissue & \begin{tabular}[c]{@{}l@{}}Acoustic\\   impedance(106Rayls)\end{tabular} \\ \hline
		Air         & 0.0004                                                                   \\
		Lung        & 0.18                                                                     \\
		Fat         & 1.34                                                                     \\
		Liver       & 1.65                                                                     \\
		Blood       & 1.65                                                                     \\
		Kidney      & 1.63                                                                     \\
		Muscle      & 1.71                                                                     \\
		Bone        & 7.8                                                                      \\ \hline
	\end{tabular}
\end{table}

\subsection{INTRODUCING THE INPUT SIGNAL OF HIFU}
To use in simulation phase in Simulink MATLAB space as shown in Fig. \ref{fig001}, to simulate HIFU waves, a saw tooth wave is used. We consider SPL, PD signal as follows \cite{hedrick2005ultrasound}:

\begin{figure}[h!]
	\includegraphics[width=\linewidth]{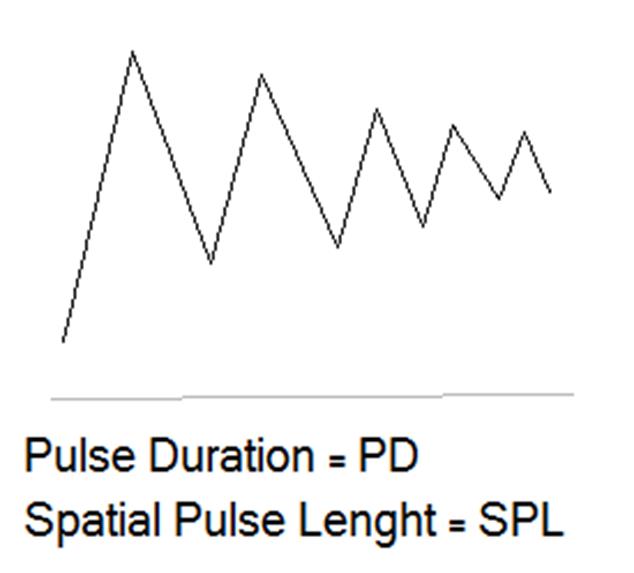}
	\caption{input wave form at the time of applying to probe HIFU for treatment}
	\label{fig001}
\end{figure}
For the pulse, SPL, PD value is subject to change based on the following values:
\begin{equation}\label{eq3}
SPL=n. \lambda 
\end{equation} 
\begin{equation}\label{eq4}
PD=n.T
\end{equation} 
In these equations SPL is adjustable according to longitudinal dimension and PD based on the dimension of time (spatial and temporal length of pulse). In these equations,  n is number of damped pulse oscillations, T, pulse period and ${\lambda}$ is the wavelength of a pulse oscillation, it means  that parameters  of  wavelength, period and number of oscillations can be changed for changing the spatial and temporal length.
If Fig. \ref{fig001} input signal as a mechanical signal enters into a layer of mass, spring and damper, will weaken sound. Weakened sound is absorbed by the mechanical tissue as thermal energy and remaining pass through it as weakened mechanical waves.
\begin{figure}[h!]
	\includegraphics[width=\linewidth]{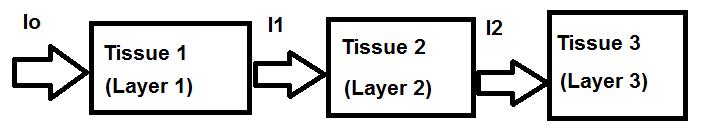}
	\caption{attenuation of sound waves from one layer to the other}
	\label{fig002}
\end{figure}
\begin{figure}[h!]
	\includegraphics[width=\linewidth]{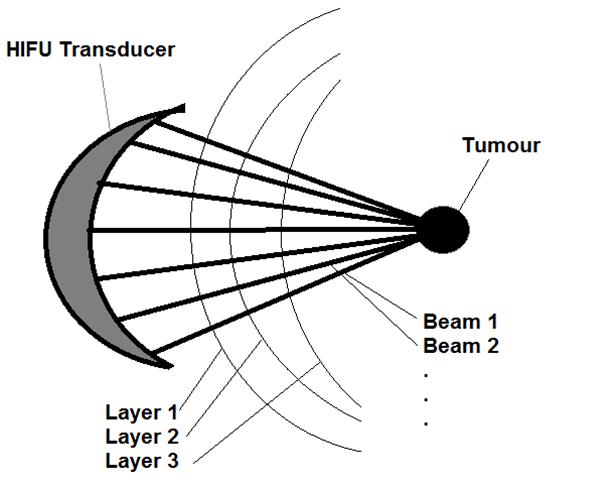}
	\caption{Ultrasound waves beam when come out of transducer and pass through three hypothetical layers}
	\label{fig003}
\end{figure}
In Fig. \ref{fig003}, when the ultrasound waves come out of HIFU transducer, become weak in each layer as shown in Fig. \ref{fig002}. Thus when ultrasound waves reach tumor tissue, have become very weak. In this time if final heat of every beam of ultrasound waves is T, in tumor tissue will be:
\\
\begin{equation}
\label{eq5}
T_{Total} = \sum_1^n  T_{n}  =  T_{1} + T_{2} + T_{3} +...
\end{equation}
\\
\begin{figure}[h!]
	\includegraphics[width=\linewidth]{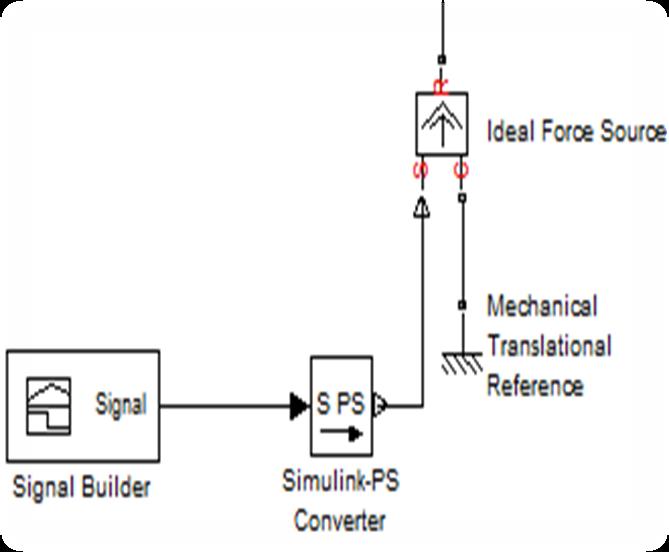}
	\caption{HIFU input electrical wave before turning to mechanical waves}
	\label{fig004}
\end{figure}
\subsection{INTRODUCING MODEL COMPONENTS}
In this model, body tissue is modelled as mechanical elements. In this model spring is considered as tissue elasticity and damper as a spring -resistant and spring fixer. The resulting force is stored as a total result in intended mass (Fig. \ref{fig004}).
According to a performed project report for HIFU simulator, it was performed by a blue phantom that target tissue was within it and obtained the heat and pressure results in the Cartesian directions \cite{jahangir2012infrared} . Accordingly, by applying HIFU wave, ISO shape of heat dose is obtained in a certain range of the Phantom. This application is also used in FDA that is as a medical device measure reference. Accordingly, it is intended to be a beam of sound waves to pass directly and reach a tissue. In simulation that was performed using SIMULINK MATLAB, three layers were considered that these three layers can be considered in line with a beam of Fig. \ref{fig003} ultrasound wave. 
If the tissue is taken into account with mechanical properties, consists of three mentioned elements. Viscoelasticity of tissue can be determined by digital signal processing works even with passing mechanical and acoustic wave \cite{eskandari2013method} . Accordingly in the following Fig layer and each layer based on the mechanical properties is considered. The temperature of each layer should be measured.  Following formula is used to do so under the terms of the heat degree electrical rules in two ends of resistance:
\begin{equation}
\label{eq6}
W=R. i^{2} 
\end{equation}
In this equation, R is the electrical resistance, I, electrical current and W is generated thermal energy. If according to the binary rules, resistance amount of a damper is supposed to be a number like b and corresponding to electric current in mechanics, sound speed is assumed v, heat amount in two ends of damper that causes heat in the model is obtained from the following equation:
\begin{equation}
\label{eq7}
W=b. v^{2} 
\end{equation}
The heat is the product of two parameters multiplication. This product can be displayed at two ends of each layer. It should be mentioned that in the simulation the time dependent variable and time varying viscosity is not assumed. Even with thermal sensors, the heat can be obtained at the ends of each layer as it is done in a separate simulation.
\begin{figure}[h!]
	\includegraphics[width=\linewidth]{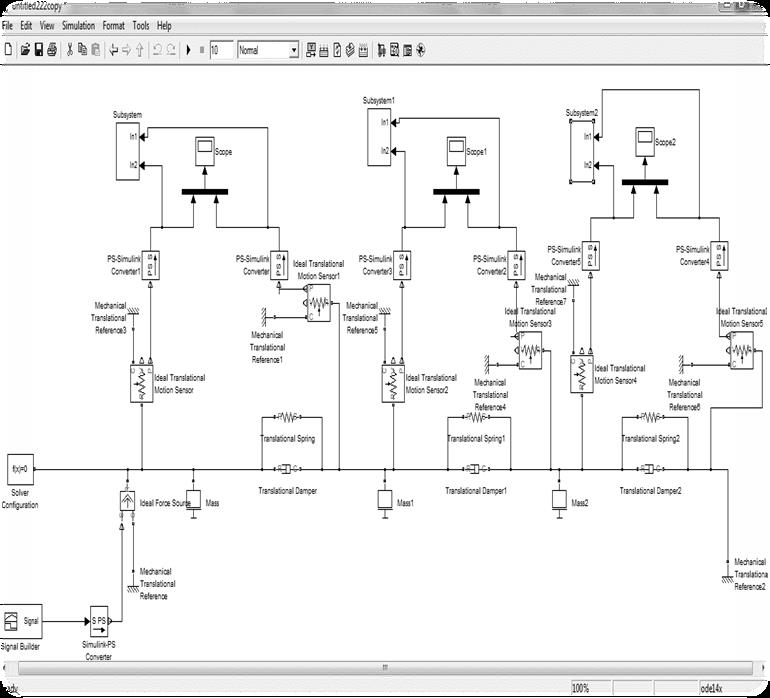}
	\caption{modeling with Simulink using sensor for obtaining pressure in three different layers}
	\label{fig005}
\end{figure}
\subsection{LIVER, KIDNEY COMPONENTS PARAMETER}
We divide a sheep kidney weighing 61.07 grams into four equal parts and the size and weight of each piece based on measurement with Verniercalliper and accurate scale is as Fig. 6, 7. Additionally the measurement system can draw strain and shear stress of one ship’s kidney same as Fig. 9 therefore we can obtain much more mechanical data of kidney.
\begin{figure}[h!]
	\includegraphics[width=\linewidth]{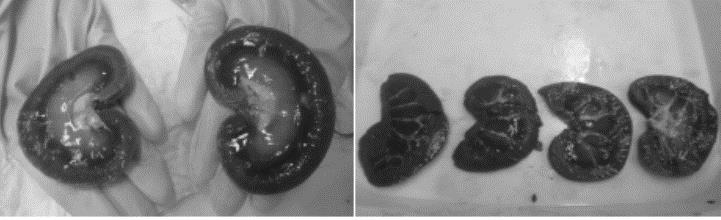}
	\caption{kidney tissue of sheep without a preservative and divided into 4 parts with cutter}
	\label{fig006}
\end{figure}
\begin{figure}[h!]
	\includegraphics[width=\linewidth]{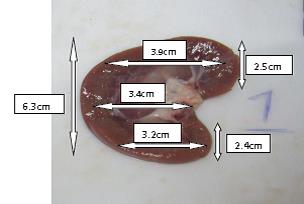}
	\includegraphics[width=\linewidth]{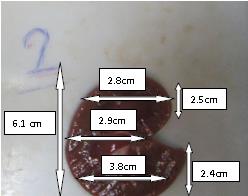}
	\caption{dimensions, volume and weight of each piece separately - part a}
	\label{fig0071}
\end{figure}[h!]
\begin{figure}[h!]
	\includegraphics[width=\linewidth]{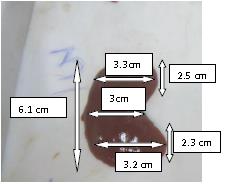}
	\includegraphics[width=\linewidth]{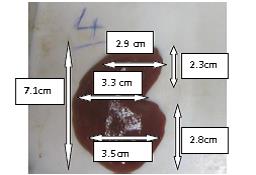}
	\caption{dimensions, volume and weight of each piece separately - part b}
	\label{fig0072}
\end{figure}
Shear stress and strain curves for a lobe symbolically with Anton Paar device is as follows. With putting the values of mass, spring, damper in Fig. \ref{fig003}, HIFU pressure on a tissue comprised of a layer of kidney can be obtained. In the phantom that is designed as follows, a water layer and a layer of tissue in the middle of it is immersed (Fig.\ref{fig008}):
\begin{figure}[h!]
	\includegraphics[width=\linewidth]{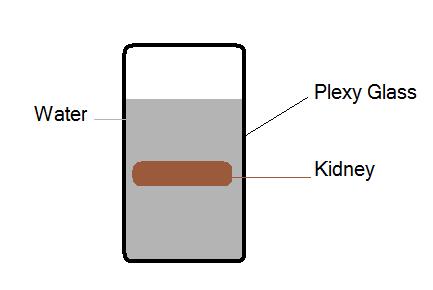}
	\caption{Assumed Phantom and with values for the treatment and simulation}
	\label{fig008}
\end{figure}
\begin{figure}[h!]
	\includegraphics[width=\linewidth]{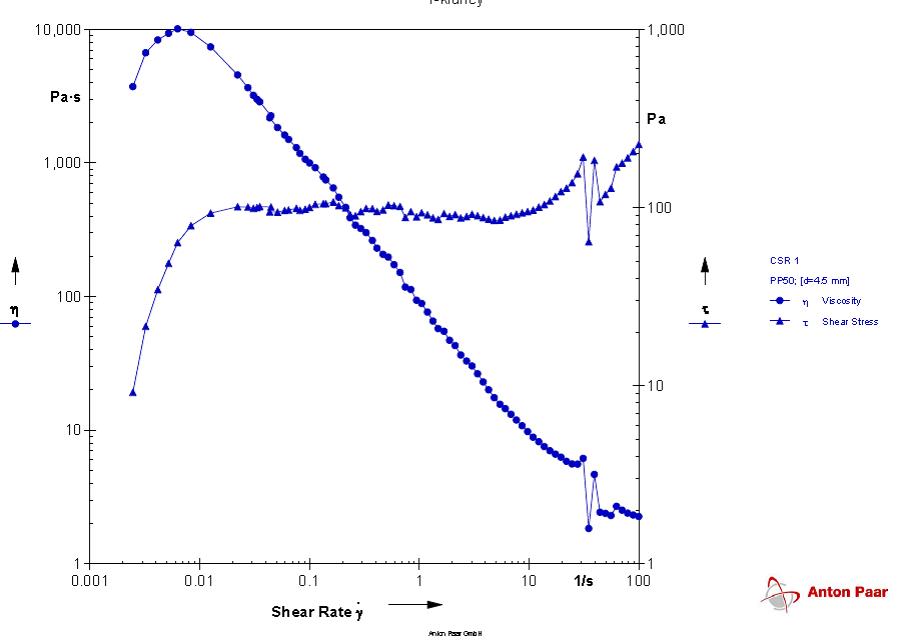}
	\caption{The chart shape of the strain and stress for a tissue layer.}
	\label{fig009}
\end{figure}
\\
\section{RESULTS}
\begin{figure}
	\includegraphics[width=\linewidth]{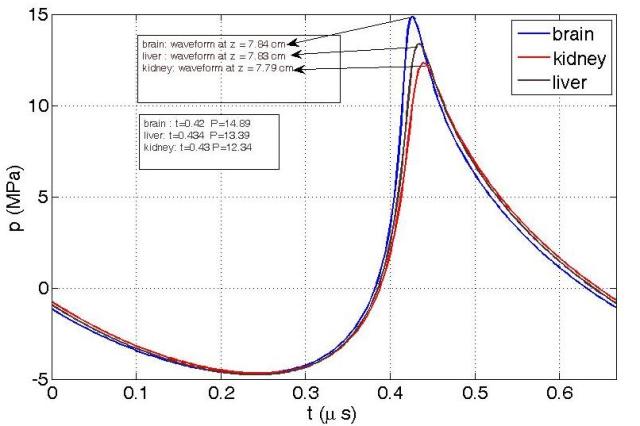}
	\includegraphics[width=\linewidth]{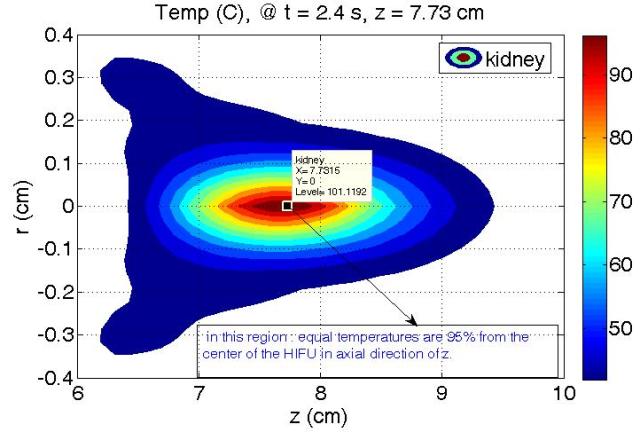}
	\caption{The generated heat shape for kidney tissue by FDA}
	\label{fig010}
\end{figure}

Using data obtained from kidney tissue with placement in Fig. \ref{fig005} simulation, amount of heat wave in three layers of kidney tissue can be obtained by applying HIFU wave .With average heat from resulting heat in the three tissues, and an average temperature in kidney tissue is estimated. Using the HIFU simulator software of FDA site, heat degree can be obtained based on FDA is official reference of standard of medical equipment according to Fig. \ref{fig010} charts by placement of acoustic impedance and sound velocity in the kidney tissue. In these comparisons this should be noted that the blood flows heat sink that makes the tissue cool is ignored and all time variable values have not been considered. 
\\
If we had access to real transducer in Iran could do more accurate comparison. For central Point of kidney tissue, heat degree was 90 $^{\circ}$ C with HIFU simulator Software and with MATLAB simulator was 79 $^{\circ}$ C which this difference in being low of the number of layers of kidney, can be lack of attention to the blood supply. 
With this simulator it is hoped that the input heat degree can be changed using the input ultrasound wave to prevent the skin lesions reported during treatment with HIFU (Fig.\ref{fig012}).
\section{DISCUSSIONS}
Look at the Fig.11 Layer model, the outermost layer of body is skin and pressure and heat leave their maximum amount on skin. Ultrasound beams are entered into the tissue from the skin. According to formula:      
\begin{equation}
\label{eq8}
I= I_{0}.exp(- \mu  * z)
\end{equation}

In this equation, μ is linear attenuation coefficient, z is depth and I is sound intensity that in the formula intensity and thickness of each layer becomes less gradually. If the pressure or temperature at the outermost layer is not calculated, leads to burning of tissue or damage of tissue. To avoid tissue damage, we should set the intensity and time of sound passes externally. For this purpose, at first the mechanical model of Fig. \ref{fig003} was used to apply according to waves tissue specifications. In this model and project, since we did not have access to a real HIFU device and a human tissue, a kidney and liver of a sheep at a later stage in Fig. \ref{fig006} phantom were used. 
In all of these steps, variable was not taken into account with time, it is   hoped that in later trials for more accuracy, variable with time was taken into account. FDA site software was used due to lack of real access. Much of this simulation was done in HFU for tissue protection and optimization of treatment planning. As ultrasound safety reports are available, ultrasound beams should always be kept in safe limit even during treatment to prevent physical and thermal damage of healthy tissue.
\begin{figure}
	\includegraphics[width=\linewidth]{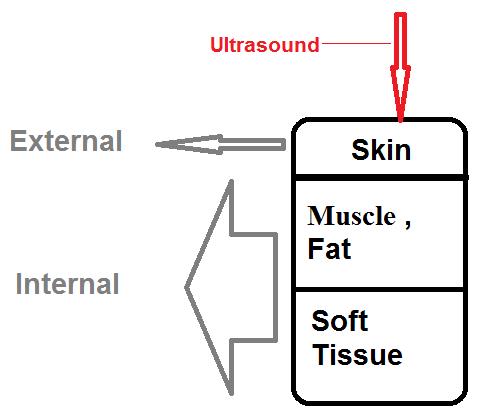}
	\caption{obtained Phantom similarity with the actual shape inside the body}
	\label{fig011}
\end{figure}
This is important when by Association of Medical Physics of America is placed in safe phase in ultrasound beam treatment phase to prevent tissue damage (Fig. 13). Example of safe environment graph of ultrasound and tissue damage that is created during treatment and non -compliance with safety standards comes in the following forms. With the completion of the modeling and having practical devices, Simulation parameters can be more complete and by obtaining the amount of heat in each layer, further tissue damage can be prevented in the future.
\begin{figure}
	\includegraphics[width=\linewidth]{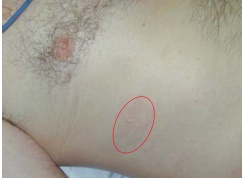}
	\caption{CTC grade 1 skin toxicity at treatment site \cite{bushberg2011essential}}
	\label{fig012}
\end{figure}
\begin{figure}
	\includegraphics[width=\linewidth]{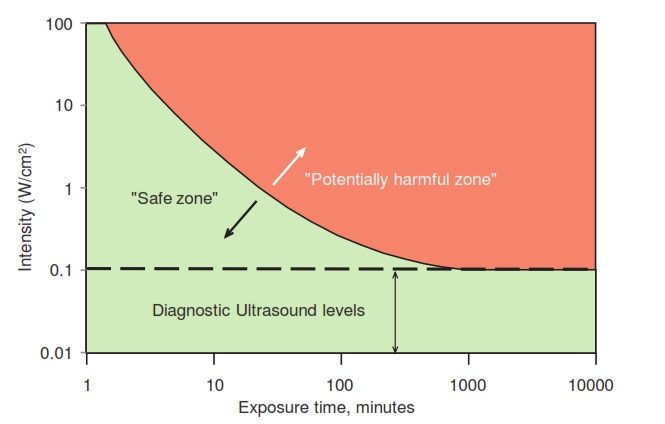}
	\caption{A diagrams of potential bio effects from ultrasound delineates safe and potentially harmful
		Regions according to ultrasound intensity levels and exposure time. The dashed line shows the upper limit of Intensities typically encountered in diagnostic imaging applications.
		}
	\label{fig013}
\end{figure}
\section{DISCUSSION AND CONCLUSION}
We used a sheep kidney tissue with definite specifications and obtained its strain and stress values after dividing it into distinct layers with specific dimensions. Those values were put in designed mechanical model then ultrasound beams were applied to it in mechanical wave form.  In this model, using temperature and binary sensors, obtained the temperature in each layer and compared it with HIFU simulation software in FDA site. Given that in the project, the weight and viscoelasticity properties of tissue have been considered and in FDA model just tissue type is considered regarding acoustic impedance and the speed of sound in tissue, comparison was made. In this comparison thermal values error in our model is about 12\% compared to the FDA error. It is possible that a lot of this error is related to simulation and comparison type error which is important and is removable in clinical trials. This comparison is considered to prevent tissue damage. In HIFU, In fact, the amount of heat in each layer reduces gradually. But phenomenon of overall effects causes a lot of pressure is applied on the tumor location or final that is attempted to consider overall work addition with this model in the next projects.
\\
\section*{ACKNOWLEDGMENT} 
We would like to thank MS Parstoo Hoseyni from Dept. of Medical Physics, Science and research branch of Islamic Azad University, Tehran, Iran, Roya Shafiyian and Sara Banayi Rad from Dept. of Biomedical Eng, Islamic Azad University, Qazvin, Iran, for their help in experimental set up and data acquisition. This work was partially supported by my father and mother and award to them.


\begin{thebibliography}{10}

\bibitem{Sarraf_2016}
S.~Sarraf and J.~Sun, ``{ADVANCES} {IN} {FUNCTIONAL} {BRAIN} {IMAGING}: A
  {COMPREHENSIVE} {SURVEY} {FOR} {ENGINEERS} {AND} {PHYSICAL} {SCIENTISTS}.,''
  {\em International Journal of Advanced Research}, vol.~4, pp.~640--660, aug
  2016.

\bibitem{aghayan2013inverse}
S.~A. Aghayan, D.~Sardari, S.~R.~M. Mahdavi, and M.~H. Zahmatkesh, ``An inverse
  problem of temperature optimization in hyperthermia by controlling the
  overall heat transfer coefficient,'' {\em Journal of Applied Mathematics},
  vol.~2013, 2013.

\bibitem{bushberg2011essential}
J.~T. Bushberg and J.~M. Boone, {\em The essential physics of medical imaging}.
\newblock Lippincott Williams \& Wilkins, 2011.

\bibitem{costa2015presage}
M.~Costa, C.~McErlean, I.~Rivens, J.~Adamovics, M.~Leach, G.~ter Haar, and
  S.~Doran, ``Presage{\textregistered} as a new calibration method for high
  intensity focused ultrasound therapy,'' in {\em Journal of Physics:
  Conference Series}, vol.~573, p.~012026, IOP Publishing, 2015.

\bibitem{eskandari2013method}
H.~Eskandari, S.~E. Salcudean, and R.~N. Rohling, ``Method and apparatus for
  determining viscoelastic parameters in tissue,'' Mar.~12 2013.
\newblock US Patent 8,394,026.

\bibitem{grady2016age}
C.~Grady, S.~Sarraf, C.~Saverino, and K.~Campbell, ``Age differences in the
  functional interactions among the default, frontoparietal control, and dorsal
  attention networks,'' {\em Neurobiology of aging}, vol.~41, pp.~159--172,
  2016.

\bibitem{guntur2015influence}
S.~R. Guntur and M.~J. Choi, ``Influence of temperature-dependent thermal
  parameters on temperature elevation of tissue exposed to high-intensity
  focused ultrasound: Numerical simulation,'' {\em Ultrasound in medicine \&
  biology}, vol.~41, no.~3, pp.~806--813, 2015.

\bibitem{hedrick2005ultrasound}
W.~R. Hedrick, D.~L. Hykes, and D.~E. Starchman, ``Ultrasound physics and
  instrumentation,'' 2005.

\bibitem{jahangir2012infrared}
M.~Jahangir~Moghadam, ``Infrared imaging method for thermographic
  characterizing of medical high intensity focused ultrasound devices,'' 2012.

\bibitem{kim2014dual}
H.~Kim and J.~H. Chang, ``Dual thermal therapeutic method for selective
  treatment of deep-lying tissue,'' in {\em 2014 IEEE International Ultrasonics
  Symposium}, pp.~21--24, IEEE, 2014.

\bibitem{kim2014mr}
E.~J. Kim, K.~Jeong, S.~J. Oh, D.~Kim, E.~H. Park, Y.~H. Lee, and J.-S. Suh,
  ``Mr thermometry analysis program for laser-or high-intensity focused
  ultrasound (hifu)-induced heating at a clinical mr scanner,'' {\em Journal of
  the Korean Physical Society}, vol.~65, no.~12, pp.~2126--2131, 2014.

\bibitem{martinez2014heat}
R.~Martinez, A.~Vera, and L.~Leija, ``Heat therapy hifu transducer electrical
  impedance modeling by using fem,'' in {\em 2014 IEEE International
  Instrumentation and Measurement Technology Conference (I2MTC) Proceedings},
  pp.~299--303, IEEE, 2014.

\bibitem{sarraf2014brain}
S.~Sarraf, C.~Saverino, H.~Ghaderi, and J.~Anderson, ``Brain network extraction
  from probabilistic ica using functional magnetic resonance images and
  advanced template matching techniques,'' in {\em Electrical and Computer
  Engineering (CCECE), 2014 IEEE 27th Canadian Conference on}, pp.~1--6, IEEE,
  2014.

\bibitem{sarraf2014mathematical}
S.~Sarraf, E.~Marzbanrad, and H.~Mobedi, ``Mathematical modeling for predicting
  betamethasone profile and burst release from in situ forming systems based on
  plga,'' in {\em Electrical and Computer Engineering (CCECE), 2014 IEEE 27th
  Canadian Conference on}, pp.~1--6, IEEE, 2014.

\bibitem{sarraf2016deepad}
S.~Sarraf, G.~Tofighi, {\em et~al.}, ``Deepad: Alzheimer′ s disease
  classification via deep convolutional neural networks using mri and fmri,''
  {\em bioRxiv}, p.~070441, 2016.

\bibitem{sarraf2016robust}
S.~Sarraf, C.~Saverino, and A.~M. Golestani, ``A robust and adaptive
  decision-making algorithm for detecting brain networks using functional mri
  within the spatial and frequency domain,'' in {\em 2016 IEEE-EMBS
  International Conference on Biomedical and Health Informatics (BHI)},
  pp.~53--56, IEEE, 2016.

\bibitem{saverino2016associative}
C.~Saverino, Z.~Fatima, S.~Sarraf, A.~Oder, S.~C. Strother, and C.~L. Grady,
  ``The associative memory deficit in aging is related to reduced selectivity
  of brain activity during encoding,'' {\em Journal of cognitive neuroscience},
  2016.

\bibitem{tamura2014visualizations}
Y.~Tamura, N.~Tsurumi, and Y.~Matsumoto, ``Visualizations of bubble motions and
  temperature rises by focused ultrasound,'' {\em Procedia Engineering},
  vol.~90, pp.~5--10, 2014.

\bibitem{wang2016simulation}
M.~Wang and Y.~Zhou, ``Simulation of non-linear acoustic field and thermal
  pattern of phased-array high-intensity focused ultrasound (hifu),'' {\em
  International Journal of Hyperthermia}, pp.~1--14, 2016.

\end{thebibliography}
\end{document}